\documentclass[aps,prb,twocolumn,showpacs,groupedaddress,10pt]{revtex4}
\usepackage{amsmath}
\usepackage{amssymb}
\usepackage{graphicx}
\usepackage{dcolumn}
\usepackage{bm}
\usepackage{color}
\usepackage{subfigure}
\usepackage{soul}

\begin{document}

\title{Quantum Monte Carlo study of the dynamic structure factor \\
in the gas and crystal phase of hard-sphere bosons}
\author{R. Rota$^{\rm a}$, F. Tramonto$^{\rm b}$, D.E. Galli$^{\rm b}$, and S. Giorgini$^{\rm a}$}
\affiliation{
 $^{\rm a}$  Dipartimento di Fisica, Universit\`a di Trento and INO-CNR BEC Center, 38123 Povo, Trento, Italy \\
$^{\rm b}$ Dipartimento di Fisica, Universit\`a degli Studi di Milano, via Celoria 16, 20133 Milano, Italy}

\begin{abstract}
We investigate the dynamic structure factor of a system of Bose particles at zero temperature using quantum Monte Carlo methods. Interactions are modeled using a hard-sphere potential of size $a$ and simulations are performed for values of the gas parameter $na^3$ ranging from the dilute regime up to densities $n$ where the thermodynamically stable phase is a solid. With increasing density we observe 
a crossover of the dispersion of elementary excitations from a Bogoliubov-like spectrum to a phonon--maxon--roton curve and the emergence of a broad multiphonon contribution accompanying the single-quasiparticle peak. In particular, for $na^3=0.2138$, which corresponds to superfluid $^4$He at equilibrium density,
the extracted spectrum turns out to be in good agreement with the experimental energy--momentum dispersion relation in the roton region and for higher momenta.
The behavior of the spectral function at the same density in the stable solid and metastable gas phase above the freezing point is also discussed.        


\end{abstract}

\maketitle

\section{Introduction}

The dynamic structure factor of a many-body system contains a wealth of information about the nature and energy spectrum of the excitations coupled to density fluctuations. At zero temperature the dynamic structure factor is defined as the Fourier transform of the density-density correlation function evaluated on the ground state $|\Psi_0\rangle$:
\begin{equation}
S({\bf q},\omega)=\frac{1}{2\pi N}\int_{-\infty}^{+\infty}dt\; e^{i\omega t} \frac{\langle\Psi_0|\rho_{-{\bf q}}(t)\rho_{\bf q}|\Psi_0\rangle}{\langle\Psi_0|\Psi_0\rangle} \;.
\label{Sqomega1}
\end{equation}
Here $\rho_{\bf q}=\sum_{i=1}^Ne^{-i{\bf q}\cdot{\bf r}_i}$, with ${\bf r}_i$ being the coordinates of the $N$ particles in the system, is the operator corresponding to a density fluctuation with wavevector ${\bf q}$ and $\rho_{\bf q}(t)=e^{iHt/\hbar}\rho_{\bf q}e^{-iHt/\hbar}$ is the same operator following a time evolution with the Hamiltonian $H$. By introducing the complete set of energy eigenstates $|\Psi_n\rangle$, the definition of $S({\bf q },\omega)$ can be equivalently expressed as the positive definite sum of terms
\begin{equation}
S({\bf q},\omega)=\frac{1}{N}\sum_{n\ge0}\delta\left(\omega-\frac{E_n-E_0}{\hbar}\right)\frac{|\langle\Psi_n|\rho_{\bf q}|\Psi_0\rangle|^2}{\langle\Psi_0|\Psi_0\rangle} \;,
\label{Sqomega2}
\end{equation}
involving all excited states with excitation energy $E_n-E_0$ from the ground state which are not othogonal to the density perturbation $\rho_{\bf q}|\Psi_0\rangle$. Important relations involving the dynamic structure factor are provided by its zero momentum
\begin{equation}
S({\bf q})=\int_0^\infty d\omega S({\bf q},\omega)=\frac{1}{N}\frac{\langle\Psi_0|\rho_{-{\bf q}}\rho_{\bf q}|\Psi_0\rangle}{\langle\Psi_0|\Psi_0\rangle} \;,
\label{Sqomega3}
\end{equation}
which defines the static structure factor $S({\bf q})$, and by its first momentum
\begin{equation}
\int_0^\infty d\omega \;\omega S({\bf q},\omega)=\frac{\hbar q^2}{2m} \;,
\label{Sqomega4}
\end{equation}
also known as $f-$sum rule and holding for any system of $N$ particles of mass $m$ interacting via velocity independent potentials.

In the context of quantum degenerate Bose systems, the dynamic structure factor has provided an invaluable tool for the experimental and theoretical investigation of superfluid $^4$He~\cite{BookGlyde} as well as of ultracold atomic gases~\cite{BookStringari}. A landmark achievement  of these studies has been the precise measurement of the phonon-maxon-roton spectrum in superfluid $^4$He and, in more recent years, important experimental contributions came from the observation of the Bogoliubov dispersion in a dilute condensate of $^{87}$Rb atoms~\cite{Steinhauer02}  and of the spin response in a strongly interacting superfluid Fermi gas~\cite{Hoinka12}.

In the most interesting regime of strong interactions, quantitative theoretical investigations of the dynamic structure factor can only rely on numerical simulations. In Ref.~\onlinecite{Boninsegni96} quantum Monte-Carlo (QMC) methods have been applied to the  calculation of $S({\bf q},\omega)$ in superfluid helium. Since this first study, similar calculations have been carried out again in systems of $^4$He atoms both in the liquid and in the solid phase~\cite{Boninsegni98, Baroni99, Vitali10, Rossi12, Roggero12}, in systems of bosons with soft-core repulsive potentials~\cite{Saccani12} and in two-dimensional systems of $^4$He\cite{Arrigoni13} and fermionic liquid $^3$He~\cite{Nava13}.
Even if equilibrium properties can be calculated exactly, the main drawback of such methods consists in the analytic continuation from purely imaginary to real time of the correlation function entering the Fourier transform in Eq.~(\ref{Sqomega1}) which strongly limits any possibility of a precise determination of $S({\bf q},\omega)$.  This is a well-known mathematically ill-conditioned problem saying that, whatever small is the statistical error in the estimate of the imaginary-time correlations, an unambiguous reconstruction of the corresponding spectral function is ruled out~\cite{Boninsegni96}. A possible strategy to partly overcome this difficulty is offered by the maximum entropy method, which has been largely utilized in the first simulations aimed to determine $S({\bf q},\omega)$~\cite{Boninsegni96, Boninsegni98, Baroni99}. More recently, an alternative method based on a genetic algorithm and known as genetic inversion via falsification of theories (GIFT) has been put forward and has already been applied in several studies~\cite{Vitali10,Rossi12,Nava13,Arrigoni13}.   

In this work we determine the dynamic structure factor at $T=0$ of a system of bosonic hard spheres, using the ``exact'' path-integral ground-state (PIGS) method to calculate the density-density correlation function in imaginary time and the GIFT method to perform the inversion in the spectral domain. The interaction parameter is varied from the weakly coupled regime, where the elementary excitations of the gas are well described in terms of Bogoliubov quasi-particles, to the regime of strong coupling where, similarly to liquid helium, the dispersion law exhibits the typical phonon-maxon-roton features. With increasing coupling, we observe the appearance of an incoherent, multiphonon contribution in the spectral function at frequencies higher than the single quasi-particle peak and we determine the density at which the dispersion curve of this peak is first featuring a roton minimum. For high densities the thermodynamically stable phase is a crystal and $S({\bf q},\omega)$ at the reciprocal lattice vectors is expected to be concentrated at zero frequency. The differences in the response function in the stable solid and in the metastable gas phase at the same density are also discussed.

The structure of the paper is as follows. In Section~\ref{Method} we introduce the model of hard-core bosons and we explain in some details the PIGS method. We also provide a precise determination of the equation of state of the
gas and solid phase close to the point of the phase transition, including the values of the freezing and melting critical densities. In Section~\ref{GIFT} we describe the calculation of the density-density correlation function in imaginary time and its inversion by means of the GIFT algorithm. In Section~\ref{Results} we report the results on $S({\bf q},\omega)$ and on the dispersion of excitations for different values of the interaction strength, both in the gas and in the solid phase. In this section we also compare both the dynamic and the static structure factor of the hard-sphere gas with the ones of superfluid $^4$He at the equilibrium density.

Finally, in the last Section, we draw our conclusions.               

\section{Method}
\label{Method}
\subsection{Quantum hard-sphere model and PIGS method}
The quantum degenerate hard-sphere (HS) gas serves as an important reference model for a many-body system with short-range repulsive interactions. The model is particularly useful in the dilute regime, where the details of the interatomic potential are irrelevant and the picture of impenetrable particles captures the essential properties of ultracold atoms with a positive scattering length~\cite{Huang57}. At higher densities the attractive tail of the potential plays a crucial role and predictions from the 
HS model can only be qualitatively correct. Nevertheless, the HS model has been used to characterize semi--quantitatively the static properties of a strongly interacting system like superfluid $^4$He~\cite{Hansen71, Kalos74}.
Moreover, the model provides one with a well defined system where quantum correlations from the weak to the strong coupling regime can be investigated by varying a single parameter, namely the reduced density in units of the HS range.

The HS model corresponds to the following Hamiltonian for $N$ identical particles of mass $m$
\begin{equation}
H=-\frac{\hbar^2}{2m}\sum_{i=1}^N\nabla_i^2+\sum_{i<j}V(|{\bf r}_i-{\bf r}_j|) \;,
\label{Hamiltonian}
\end{equation} 
with
\begin{eqnarray}\label{hardspherepotential}
V(r) = \left\{ 
\begin{array}{cll} \infty           & (r \le a)  \\
                      0             & (r>a) \;.
\end{array}\right.
\label{HSpot}
\end{eqnarray}
We notice that the range $a$ of the HS potential coincides with the $s$-wave scattering length of the two-body problem. The model contains a single energy scale $\hbar^2/(2ma^2)$ and a single dimensionless parameter (gas parameter) $na^3$, where $n=N/V$ is the particle density. The system volume $V=L^3$ consists of a cubic box of size $L$ and periodic boundary conditions are enforced to simulate the infinite system.

In the dilute regime ($na^3\ll1$) the equation of state of the HS gas at zero temperature has been established using perturbation theory in a series of fundamental papers~\cite{Huang57} for both bosonic and fermionic particles. Results for the ground-state energy of a Bose HS system have been obtained over a wide range of densities using exact quantum Monte Carlo methods~\cite{Kalos74,Giorgini99}. Similar results for a two-component Fermi system with HS interspecies interactions were reported in Refs.~\onlinecite{Pilati10, Chang10} from calculations based on the fixed-node diffusion Monte Carlo method. 

At very high density the ground state of a HS system should be a fcc crystal similarly to the classical case~\cite{Hoover68, Woodcock97}. The freezing ($n_f$) and melting ($n_m$) densities have been determined for Bose particles at $T=0$ using the variational Monte Carlo method~\cite{Hansen71} [$n_fa^3=0.23(2)$ and $n_ma^3=0.25(2)$], the Green's function Monte Carlo method~\cite{Kalos74} [$n_fa^3=0.25(1)$ and $n_ma^3=0.27(1)$] and the density functional approach~\cite{Denton90} [$n_fa^3=0.25$ and $n_ma^3=0.28$].

In this work, the properties of a Bose HS system at zero temperature are calculated using an exact numerical technique: the PIGS method ~\cite{Sarsa00}. In contrast to the diffusion Monte Carlo method, the PIGS method allows for an unbiased estimate also of non-local observables such as the one-body density matrix (OBDM), whose off-diagonal terms at large distance define the fraction of condensed particles in the system~\cite{Tramonto08,Rossi09}. 

The PIGS method is a projection technique in imaginary time that, starting from an initial wavefunction $\Psi_T({\bf R})$, where ${\bf R}=({\bf r}_1,\dots,{\bf r}_N)$ denotes the set of spatial coordinates of the $N$ particles, projects it onto the ground-state wavefunction $\Psi_0({\bf R})$ after evolution over a long enough imaginary time interval $\tau$. In fact, if $\Psi_T({\bf R})$ is not orthogonal to the ground state, the following relation holds
\begin{eqnarray}
\Psi_0({\bf R})&=&{\cal{N}}\lim_{\tau\to\infty}e^{-\tau H}\Psi_T({\bf R})
\nonumber\\
&=&{\cal{N}}\lim_{\tau\to\infty}\int d{\bf R}^\prime 
G({\bf R},{\bf R}^\prime,\tau)\Psi_T({\bf R}^\prime) \;,
\label{PIGS1}
\end{eqnarray}  
where ${\cal{N}}$ is a normalization factor and in the second equation we introduced the Green's function $G({\bf R},{\bf R}^\prime,\tau)=\langle{\bf R}|e^{-\tau H}|{\bf R}^\prime\rangle$. The time propagator is in general not known but suitable approximation schemes are available at small time steps $\delta\tau$. Using one of these approximation schemes in place of the true Green's function, $G({\bf R},{\bf R}^\prime,\delta\tau)$, is one of the two approximations which characterizes the PIGS method. Then, the convolution formula allows one to calculate $G({\bf R},{\bf R}^\prime,\tau)$ using a large enough number $M-1$ of intermediate points such that $M=\tau/\delta\tau$
\begin{equation}
G({\bf R}_M,{\bf R}_0,\tau)=\int d{\bf R}_1... d{\bf R}_{M-1}\prod_{j=0}^{M-1}
G({\bf R}_{j+1},{\bf R}_j,\delta\tau) \;.
\label{PIGS2}
\end{equation} 
Then the ground-state wavefunction $\Psi_0({\bf R})$ can be approximated with the multidimensional integral
\begin{equation}
\Psi_0({\bf R}_M) \simeq \int \prod_{j=0}^{M-1} d{\bf R}_j\;
G({\bf R}_{j+1},{\bf R}_j,\delta\tau)\Psi_T({\bf R}_0) \;.
\label{PIGS3}
\end{equation}
This choice corresponds to limit the imaginary time propagation to $\tau=M\delta\tau$; this is the second approximation
used with the PIGS method. The method is ``exact'' whenever this and the previous approximation affect the computed
expectation values to an extent which is below their statistical uncertainty; such regime is always attainable
by taking $M$ large enough and $\delta\tau$ small enough.

The expectation value of a local observable, corresponding to the operator $O$, can be obtained from an average over $2M+1$ configuration points 
\begin{equation}
\frac{\langle\Psi_0|O|\Psi_0\rangle}{\langle\Psi_0|\Psi_0\rangle} = \int d{\bf R}_0...d{\bf R}_{2M}
\;p({\bf R}_0,\dots,{\bf R}_{2M}) O({\bf R}_M) \;,
\label{PIGS4}
\end{equation}
sampled from the probability distribution
\begin{eqnarray}
&&p({\bf R}_0,...,{\bf R}_{2M})=
\label{PIGS5}\\
&=&\frac{\prod_{j=0}^{2M}\Psi_T({\bf R}_{2M})G({\bf R}_{j+1},{\bf R}_j,\delta\tau)\Psi_T({\bf R}_0)}{\int \prod_{j=0}^{2M} d{\bf R}_j \Psi_T({\bf R}_{2M}) G({\bf R}_{j+1},{\bf R}_j,\delta\tau)\Psi_T({\bf R}_0)}\;.
\nonumber
\end{eqnarray}
Estimates like Eq.~(\ref{PIGS4}) can be interpreted as ensemble averages with the probability distribution (\ref{PIGS5}) over a classical system of $N$ open polymers.  Each polymer is formed by $2M+1$ monomers (beads) corresponding to the coordinates $({\bf r}_i^{(0)},...,{\bf r}_i^{(2M)})$ which determine the points ${\bf R}$ in configuration space: ${\bf R}_j=({\bf r}_1^{(j)},...,{\bf r}_N^{(j)})$. By virtue of the definition of $\Psi_0$ in Eq.~(\ref{PIGS3}), only the central beads of the polymers are sampled according to the square of the ground-state wave function and thus local operators, such as the one entering Eq.~(\ref{PIGS4}), are evaluated only at the mid-point configuration ${\bf R}_M$. If we deal with non-local observables, such as the kinetic energy or any correlation function in imaginary time, Eq.~(\ref{PIGS4}) has to be modified since the corresponding estimators depend on more configuration points ${\bf R}$. Nonetheless, it is important to choose a sufficiently large number of projections $M$ and to consider only the central part of the classical polymers for the evaluation of the averages. In the particular case of the ground-state energy, the identity $\langle\Psi_0|H|\Psi_0\rangle=\langle\Psi_T|e^{-2\tau H}H|\Psi_T\rangle$, holding if $\tau$ is large enough, provides one with an estimate in terms of the local energy $E_L({\bf R})=\Psi_T^{-1}({\bf R})H\Psi_T({\bf R})$, which is conveniently calculated starting from the trial wavefunction and has to be evaluated on either of the configurations, ${\bf R}_0$ or ${\bf R}_{2M}$, at the end-point of the polymers. Alternative estimators of the energy are also available, such as the direct and the virial estimator~\cite{Cuervo05,Ceperley95}. However, unless the trial wavefunction is a particularly poor approximation of the ground-state wavefunction, it is preferable to use the local energy estimator for the calculation of $\langle\Psi_0|H|\Psi_0\rangle$, since the other estimators tipically suffer of a larger variance.

In homogeneous systems the OBDM is defined via the equation $\rho_1(s)=\frac{\langle\Psi_0|\psi^\dagger({\bf r}+{\bf s})\psi({\bf r})|\Psi_0\rangle}{\langle\Psi_0|\Psi_0\rangle}$, in terms of the annihilation and creation operators of particles. The first quantization expression of $\rho_1(s)$ is given by
\begin{equation}
\rho_1(s)=\frac{\int d{\bf r}_2...d{\bf r}_N \Psi_0^\ast({\bf r}+{\bf s},...,{\bf r}_N)
\Psi_0({\bf r},...,{\bf r}_N)}{\int d{\bf r}_1...d{\bf r}_N |\Psi_0({\bf r}_1,...,{\bf r}_N)|^2} \;,
\label{PIGS6}
\end{equation}
and is suited for an estimate with the PIGS method obtained by cutting one of the $N$ polymers at the position of the $M$-th bead, i.e. dividing one polymer into two half polymers with $M+1$ beads each, and by collecting the occurrences of the distances between the two loose ends. An efficient sampling of the relevant configurations and the normalization of $\rho_1$ are provided by the worm algorithm, a technique developed for path integral Monte Carlo simulations at finite temperature~\cite{Boninsegni06} in which the configurations are allowed to switch from the ``diagonal'' sector (all $N$ polymers with $2M+1$ beads) to the ``off-diagonal" sector ($N-1$ polymers with $2M+1$ beads and two halves polymers with $M+1$ beads each) and viceversa. Swap moves are also allowed in the ``off-diagonal'' sector, where one full and one half polymer are exchanged by cutting the full polymer into two halves and by merging one of the two with the half polymer. These latter moves enforce the proper sampling of particle permutations.   

The evaluation of the Green's function at small time steps can be performed, in analogy with the path-integral Monte Carlo method at finite temperauture, using the pair-action approximation~\cite{Pollock84, Ceperley95}
\begin{equation}
G({\bf R},{\bf R}^\prime,\delta\tau)=G_0({\bf R},{\bf R}^\prime,\delta\tau)
\prod_{i<j}\frac{G_{rel}({\bf r}_{ij},{\bf r}_{ij}^\prime,\delta\tau)}{G_{rel}^0({\bf r}_{ij},{\bf r}_{ij}^\prime,\delta\tau)} \;.
\label{PAIR1}
\end{equation}
Here $G_0$ is the free-particle propagator consisting in the contribution of the kinetic energy, $T=-(\hbar^2/2m)\sum_{i=1}^N\nabla_i^2$, to the Green's function:
\begin{eqnarray}
G_0({\bf R},{\bf R}^\prime,\delta\tau)&=&\langle{\bf R}|e^{-\delta\tau T}|{\bf R}^\prime\rangle
\label{PAIR2}\\
&=&\prod_{i=1}^N\left(\frac{m}{2\pi\hbar^2\delta\tau}\right)^{3/2}e^{-m({\bf r}_i-{\bf r}_i^\prime)^2/
(2\hbar^2\delta\tau)} \;.
\nonumber
\end{eqnarray}
The function $G_{rel}$ is the two-body propagator of the interacting system, which depends on the relative coordinates ${\bf r}_{ij}={\bf r}_i-{\bf r}_j$ and ${\bf r}_{ij}^\prime={\bf r}_i^\prime-{\bf r}_j^\prime$, and in Eq.~(\ref{PAIR1}) it is divided by the corresponding non-interacting term
\begin{equation}
G_{rel}^0({\bf r}_{ij},{\bf r}_{ij}^\prime,\delta\tau)=
\left(\frac{m}{4\pi\hbar^2\delta\tau}\right)^{3/2}e^{-m({\bf r}_{ij}-{\bf r}_{ij}^\prime)^2/
(4\hbar^2\delta\tau)} \;. 
\label{PAIR3}
\end{equation}
In the case of the HS potential a high energy expansion of the two-body propagator due to Cao and Berne~\cite{CaoBerne} provides an accurate approximation holding at small time steps
\begin{eqnarray}
\frac{G_{rel}({\bf r}_{ij},{\bf r}_{ij}^\prime,\delta\tau)}{G_{rel}^0({\bf r}_{ij},{\bf r}_{ij}^\prime,\delta\tau)}
&=&1-\frac{a(r+r^\prime)-a^2}{rr^\prime}
\label{PAIR4}\\
&\times& e^{-m[rr^\prime+a^2-a(r+r^\prime)](1+\cos\theta)/(2\hbar^2\delta\tau)} \;,
\nonumber
\end{eqnarray}
where $\theta$ is the angle between ${\bf r}$ and ${\bf r}^\prime$.

A further ingredient of the method is provided by the trial function $\Psi_T({\bf R})$. For the gas branch we use the translationally invariant Jastrow function 
\begin{equation}
\Psi_T({\bf R})=\prod_{i<j} f(|{\bf r}_i-{\bf r}_j|) \;,
\label{WF1}
\end{equation}
where the two-body correlation function $f$ is chosen as $f(r)=\sin(k(r-a))/r$ if $r>a$ and $f(r)=0$ if $r<a$.
This choice corresponds to the two-body $s$-wave scattering solution from a HS potential and the wavevector $k$ is chosen such as the derivative $f^\prime(r)$ vanishes at $r=L/2$ to fulfill periodic boundary conditions. To simulate the solid we instead use a wave function that explicitly breaks translational symmetry, obtained by multiplying the Jastrow function (\ref{WF1}) by a one-body term which localizes the particles at the lattice sites of the crystal:
\begin{equation}
\Psi_T({\bf R})=\prod_{i=1}^Ne^{-({\bf r}_i-{\bf S}_i)^2/\alpha^2}\prod_{i<j} f(|{\bf r}_i-{\bf r}_j|) \;.
\label{WF2}
\end{equation}
Here, the localization term is a gaussian whose width $\alpha$ is a variational parameter and the lattice sites $({\bf S}_1,\dots,{\bf S}_N)$ correspond to a fcc crystal. We notice that the above wave function is not symmetric under particle exchange. Still, permutations are correctly sampled during the simulation by means of the swap moves in the worm algorithm, resulting in a symmetric ground-state wave function ~\cite{Tramonto08,Rossi09} $\Psi_0({\bf R}_M)$ as obtained from Eq.~(\ref{PIGS3}). The proper symmetry of $\Psi_0({\bf R}_M)$ is crucial when calculating the OBDM.

\begin{figure}[t]
\includegraphics[angle=-90,width=9.0cm]{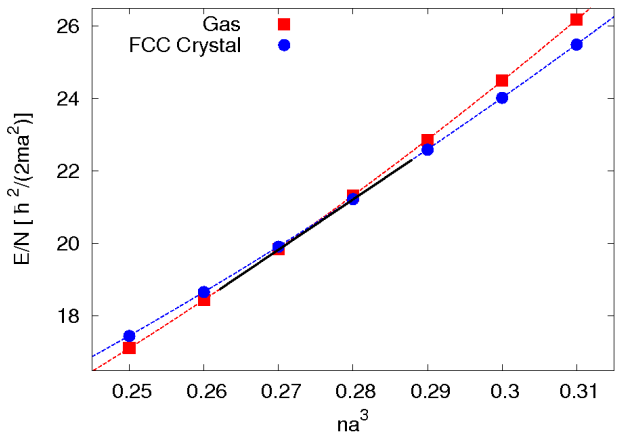}
\caption{(color online). Equation of state of the gas (red symbols) and of the fcc crystal (blue symbols) as a function of the gas parameter $na^3$. Dashed lines are polynomial fits to the PIGS results for the two phases: for the gas phase, the PIGS data are fitted with the equation $E_G = A_G (na^3 - \rho_G)^2 + E_{0,G}$, with $A_G = 357.5 \pm 5.1 $, $\rho_G = 0.0686 \pm 0.0029$, $E_{0,G} = 5.35 \pm 0.21$; for the solid phase, the PIGS data are fitted with the equation $E_S = A_S (na^3 - \rho_S)^2 + E_{0,S}$, with $A_S = 272.7 \pm 6.0$, $\rho_S = 0.0342 \pm 0.0054$, $E_{0,S} = 4.76 \pm 0.37$ (the values of $A_G$, $E_{0,G}$, $A_S$ and $E_{0,S}$ are in units of $\hbar^2/(2 m a^2)$, the parameters $\rho_G$ and $\rho_S$ are dimensionless). The black solid line corresponds to the double tangent construction and its low and high density ending points indicate respectively the values of the freezing ($n_f$) and melting ($n_m$) densities.}
\label{figure1}
\end{figure}

Simulations are carried out starting from configurations distributed according to the gas and solid wave functions, respectively Eq.~(\ref{WF1}) and (\ref{WF2}). Different system sizes are simulated: up to $N=300$ in the gas and to $N=500$ in the solid, and the corresponding energies are extrapolated to the thermodynamic limit using a linear $1/N$ fit. The results for the energy of the two phases, in the region close to the gas--to--solid phase transition, are shown in Fig.~\ref{figure1}. The PIGS method has been shown to be able to generate the correct ground state of the system irrespective of the starting configuration and of the trial wave function utilized~\cite{Rossi09,Rota10}. However, for values of the gas parameter close to the gas--to--solid transition, we are able to stabilize the metastable solid and gas phase (see Fig.~\ref{figure1}) thanks to a proper choice of the initial configuration and of the number of beads $M$. From the value of the static structure factor $S({\bf G})$ calculated at the reciprocal lattice vectors ${\bf G}$, we check that the configurations obtained from the PIGS method are belonging to the solid or to the gas branch. By applying the Maxwell construction to the equation of state of the gas and of the solid we accurately determine the freezing, $n_fa^3=0.262(1)$, and melting densities, $n_ma^3=0.288(1)$, which are in agreement with previous findings~\cite{Hansen71,Kalos74,Denton90}. Recently, the freezing and the melting densities of the HS system have been determined in another PIGS calculation\cite{Rossi13}: the melting density calculated in this work agrees with our result, whereas a small discrepancy is found in the value of the freezing density. We remark that, in the range of densities studied, the energies of the bcc and hcp crystal are found to coincide, within our statistical uncertainty, with the ones reported in Fig.~\ref{figure1} for the fcc solid.

\begin{figure}[t]
\includegraphics[angle=-90,width=9.0cm]{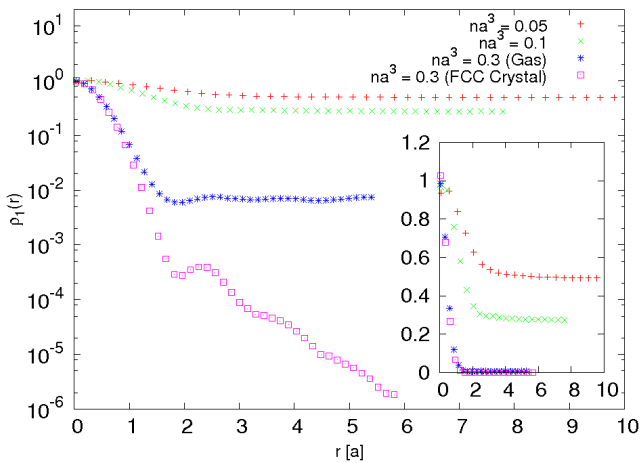}
\caption{(color online). Radial dependence of the OBDM for different values of the gas parameter in the gas and in the solid phase. The same curves are shown in logarithmic scale (main figure) and in linear scale (inset).}
\label{figure2}
\end{figure}

In Fig.~\ref{figure2} we show data corresponding to the OBDM both in the gas and in the solid phase. In the gas, the plateau reached at large distances corresponds to the condensate fraction $n_0=N_0/N$, where $N_0$ is the number of particles occupying the ${\bf k}=0$ single-particle state. We notice that $n_0$ decreases by increasing the density, but remains finite in the gas branch even in the metastable region. On the contrary, in the solid the condensate fraction vanishes, showing that off-diagonal long-range order in the OBDM does not survive when translational symmetry is broken in the HS system. It has been recently shown that in repulsive models of bosons with a soft core, $n_0$ can remain finite in the crystal phase realizing the so called supersolid state~\cite{Cinti10,Saccani11}.

\subsection{Dynamic structure factor and GIFT algorithm}
\label{GIFT}
The direct output of the PIGS algorithm is the following correlation function in imaginary time 
\begin{equation}
F({\bf q},\tau) = \frac{1}{N}\frac{\langle\Psi_0|e^{\tau H}\rho_{-{\bf q}}e^{-\tau H}\rho_{\bf q}|\Psi_0
\rangle}{\langle\Psi_0|\Psi_0\rangle} \;.
\label{GIFT1}
\end{equation} 
If the projection time onto the ground-state wave function is much longer than the time difference appearing
in $F({\bf q},\tau)$, namely $M\delta\tau\gg\tau$, the above correlation function can be calculated in terms
of the probability distribution defined in Eq.~(\ref{PIGS5})
\begin{eqnarray}
F({\bf q},\tau)=\frac{1}{N} \int d{\bf R}_0\dots d{\bf R}_{2M} \;p({\bf R}_0,\dots,{\bf R}_{2M}) &&
\nonumber\\
\times\rho_{-{\bf q}}({\bf R}_{M+\tau/\delta\tau})\rho_{\bf q}({\bf R}_M)\;, &&
\label{GIFT2}
\end{eqnarray} 
where $\rho_{\bf q}({\bf R}_j)=\sum_{i=1}^Ne^{i{\bf q}\cdot{\bf r}_i^{(j)}}$ denotes the density fluctuation operator corresponding to the point in configuration space ${\bf R}_j=({\bf r}_1^{(j)},\dots,{\bf r}_N^{(j)})$. We notice that the $\tau=0$ value of the scattering function (\ref{GIFT1}) coincides with the static structure factor of Eq.~(\ref{Sqomega3})
\begin{equation}
F({\bf q},\tau=0)=S({\bf q}) \;.
\label{GIFT3}
\end{equation}
For finite values of $\tau$, $F({\bf q},\tau)$ is instead related to the dynamic structure factor (\ref{Sqomega1}) by the Laplace transform
\begin{equation}
F({\bf q},\tau)=\int_0^\infty \!\! d\omega\; e^{-\omega\tau} S({\bf q},\omega)  \;.
\label{GIFT4}
\end{equation}
This is the equation which should be inverted in order to determine $S({\bf q},\omega)$ from the 
imaginary--time intermediate scattering function $F({\bf q},\tau)$. As we already mentioned, the problem is ill conditioned since many very different functions $S({\bf q},\omega)$, spanning from featureless to rich in structure spectral functions, are compatible with the QMC results of $F({\bf q},\tau)$ for whatever small statistical uncertainty is associated to these results.  

Different strategies have been used to extract real-time response from imaginary-time correlation functions
obtained by means of QMC calculations.
The maximum entropy method has been first applied to lattice models~\cite{Silver90,Jarrell96} and later to
continuous systems such as liquid $^4$He~\cite{Boninsegni96}.
The method uses probability theory to infer
the most probable $S({\bf q},\omega)$ by minimizing the $\chi^2$ measure for the quality of the fit of 
$F({\bf q},\tau)$, calculated from Eq.~(\ref{GIFT4}), to the QMC data and by maximizing an entropy 
functional which embodies some prior information about $S({\bf q},\omega)$, such as its positive definiteness
and the fulfillment of sum rules.

The recently introduced GIFT method offers an alternative approach: given the large number of dynamic
structure factors $S({\bf q},\omega)$ compatible with the QMC estimations of $F({\bf q},\tau)$ via \eqref{GIFT4},
the aim of the GIFT
method is to collect a large group of such spectral functions in order to discern the presence of common features,
(like, for example, support, peaks positions and intensities) in the majority of them.
The idea\cite{Tarantola06} underlying this procedure is that only such common features, shared by the majority of the
spectral functions compatible with $F({\bf q},\tau)$, can be ascribed to the true dynamic structure factor. 

As discussed above, QMC projector techniques are based on
a discretization of the imaginary time domain, with time step $\delta\tau$.
For a given wavevector ${\bf q}$, such discretization allows for
an estimation of $F({\bf q},\tau)$ only in correspondence
with a finite number of imaginary time values $\{0,\delta\tau,2\delta\tau,...,l\delta\tau\}$,
$\mathcal{F} \equiv \{F({\bf q},0), F({\bf q},\delta\tau), ..., F({\bf q},l\delta\tau)\}$.
In general $\mathcal{F}$ is obtained as an average of several
QMC calculations of $F({\bf q},\tau)$, each affected by statistical noise and which are used to estimate the
{\it statistical uncertainties} $\{\sigma_0,\sigma_1,...,\sigma_l\}$ associated with $\mathcal{F}$.
The task is then to evaluate the dynamic structure factor starting from limited and noisy data.
Often sum rules provide useful help, either imposing exact constraints on
$S({\bf q},\omega)$ or allowing to perform additional QMC measurements
which provide estimations for some moments of
$S({\bf q},\omega)$:
$\mathcal{C}\equiv \{c_n=\int_{-\infty}^{+\infty} d\omega \omega^n S({\bf q},\omega),\,n\in\mathbb{Z}\}$.
Moreover some {\it a priori knowledge} may be assumed such as the support,
non--negativity or some further properties.
When $\mathcal{F}$ has large statistical uncertainties, too much different spectral
functions would result compatible to $\mathcal{D}=\{\mathcal{F},\mathcal{C}\}$ and no common features
could be identified; on the contrary, in presence of
high accuracy in the QMC estimation of $F({\bf q},\tau)$ one could expect that some relevant features
of the true $S({\bf q},\omega)$ might be shared by the majority of the spectral functions compatible with
$\mathcal{F}$, via \eqref{GIFT4}, and $\mathcal{C}$.

We now introduce how the GIFT method concretely implements such statistical inversion scheme.
The GIFT algorithm needs (i) a space of models $\mathcal{S}$, containing a wide collection
of spectral functions consistent with any {\it prior knowledge} about $S({\bf q},\omega)$,
(ii) a {\em falsification} procedure relying on the QMC ``measurements'' $\mathcal{D}=\{\mathcal{F},\mathcal{C}\}$ and
(iii) a simple strategy to capture the accessible physical properties of the true $S({\bf q},\omega)$.

Let $\{s_1, s_2, ..., s_{N_\omega} \}$ be a set of $N_\omega$ non--negative integers ({\em spectral weights}),
whose sum is a value $\mathcal{M}$.
We partition an interval of the real positive axis $\lbrack 0,\omega_{max}\rbrack$, which we assume to be 
greater than the support of $S({\bf q},\omega)$, in subintervals all of width $\Delta\omega=\omega_{max}/N_\omega$. 
The space of models, $\mathcal{S}$, is made of linear combinations of delta functions,
i.e. we rely on models $\overline{S}$ of the form:
\begin{equation}
\label{solutions2}
\overline{S}({\bf q},\omega) = \sum_{j=1}^{N_\omega}\frac{s_j}{\mathcal{M}\Delta\omega}\delta(\omega-\omega_j),
\quad \sum_{j=1}^{N_\omega} s_j = \mathcal{M} \quad ,
\end{equation}
where $\omega_j$ represents the middle points of the $N_\omega$ subintervals of $\lbrack 0,\omega_{max} \rbrack$
and $\mathcal{M}$ provides the maximum number of quanta of spectral weight available
for the ensemble of the intervals.
From Eq.~(\ref{solutions2}) it is evident that $\int_0^\infty \overline{S}({\bf q},\omega) = 1 \;\; \forall \; {\bf q}$,
thus $\overline{S}({\bf q},\omega)$ differs from
the physical dynamic structure function $S({\bf q},\omega)$ by a factor $c_0$,
the zero--momentum, which belongs to the set of observations.

We explore $\mathcal{S}$ and check the compatibility of its elements with QMC observations,
$\mathcal{D}$, with a genetic algorithm; genetic algorithms provide an extremely efficient tool 
to explore a sample space
by a non-local stochastic dynamics, via a survival--to--compatibility evolutionary process
mimicking the natural selection rules; such evolution aims toward an increasing compatibility
with $\mathcal{D}$.
Taking into account the estimated statistical noise of $\mathcal{D}$,
any set $\mathcal{D}^{\star}$ compatible with $\mathcal{D}$ provides equivalent information
to build a compatibility function.
Thus in our genetic algorithm any random set $\mathcal{D}^{\star}=\{\mathcal{F}^{\star},\mathcal{C}^{\star}\}$
obtained by sampling independent Gaussian distributions centered on the original observations $\mathcal{D}$,
with variances which correspond to the estimated statistical uncertainties,
can be used to define the {\em compatibility}:
\begin{eqnarray}
\Phi_{\mathcal{D}^{\star}}(\overline{S}) &=&
-\sum_{j=0}^l \left[ F^{\star}({\bf q}, j \delta\tau) - \int d\omega \, e^{-\omega j \delta\tau}
c_0^{\star}\,\overline{S}({\bf q},\omega)\right]^2 \nonumber
\\
&-& \sum_n \gamma_n \left[c_n^{\star} - \int d\omega \,\omega^n
\,c_0^{\star}\,\overline{S}({\bf q},\omega)\right]^2
\label{fitness}
\end{eqnarray}
where the free parameters $\gamma_n > 0$ are adjusted in order to make the
contributions to $\Phi_{\mathcal{D}^{\star}}$ coming from $\mathcal{F}^{\star}$ and from
$\mathcal{C}^{\star}$ of the same order of magnitude.
If it happens that one $c_n$ is exactly known, no error is added making $c_n^{\star}=c_n$.
In our genetic algorithm, we start randomly constructing a collection of $\overline{S}({\bf q},\omega)$;
each $\overline{S}({\bf q},\omega)$ is coded by $N_\omega$ integers, $s_j$ in equation \eqref{solutions2}.
The genetic dynamics then consists in a succession
of {\it generations} during which an initial {\em population} of spectral functions is
replaced with new ones in order to reach regions of $\mathcal{S}$
where high values of $\Phi_{\mathcal{D}^{\star}}(\overline{S})$ exist, for a given $\mathcal{D}^{\star}$.
In the passage between two generations a succession of ``biological--like'' processes take place:
namely {\it selection}, {\it crossover} and {\it mutation}, which
are operators acting on $\mathcal{S}$ devised in such
a way to comply with the definition in equation \eqref{solutions2}.
Technical details of the method and of the implemented
genetic algorithm can be found in Ref.~\onlinecite{Vitali10}.

The GIFT method is thus based on a genetic algorithm
to propose new trials $S({\bf q},\omega)$, satisfying a number of consistency constraints, which should be
tested against the QMC data.
In our context the genetic algorithm dynamics performs a {\em falsification procedure}:
only the $\overline{S}({\bf q},\omega)$ with the highest compatibility in the last generation provides
a model for $S({\bf q},\omega)$ which has not been falsified by $\mathcal{D}^{\star}$.
Many independent evolutionary processes, say $N_{\overline{S}}$, may be generated by sampling different
$\mathcal{D}^{\star}$, thus obtaining the set $\mathcal{S}_{\mathcal{D}^{\star}}$ made
of the $N_{\overline{S}}$ different
elements $c_0^{\star}\,\overline{S}({\bf q},\omega)$;
at this point an averaging procedure inside $\mathcal{S}_{\mathcal{D}^{\star}}$
appears as the most natural way to extract shared features, i.e. relevant physical information:
\begin{equation}
S_{GIFT}({\bf q},\omega) = \frac{1}{N_{\overline{S}}} \sum_{j=1}^{N_{\overline{S}}} c_{0,j}^{\star} \; \overline{S}_j({\bf q},\omega)
\quad .
\end{equation}
In fact, in $S_{GIFT}({\bf q},\omega)$ only those features present in the majority of the spectral functions
inside $\mathcal{S}_{\mathcal{D}^{\star}}$ will survive to the average procedure, whereas undetermined
features will be averaged out.
When applied to Helium liquids, the GIFT method has provided very accurate results\cite{Vitali10,Rossi12,Nava13,Arrigoni13},
recovering, in the $^4$He case, spectral functions with sharp single quasi--particle excitations and
separating quantitatively the single quasi--particle excitation peak from the multiphonon contributions.

\section{Results}
\label{Results}
\subsection{Gas phase}
The calculation of the dynamic structure factor in a HS Bose system at a given gas parameter $na^3$ has been carried out by performing PIGS simulations with $N = 400$ particles interacting with the two-body potential in Eq.~(\ref{hardspherepotential}) and confined in a cubic box with periodic boundary conditions. The optimal values of the imaginary time step $\delta\tau$ and of the number of beads $M$, which both depend on the gas parameter, have been chosen studying the convergence of the mean energy per particle for small time step $\delta\tau$ and for large evolution time $\tau$.

\begin{figure}
\centering
\subfigure[]{
\includegraphics[angle=-90,width=7.5cm]{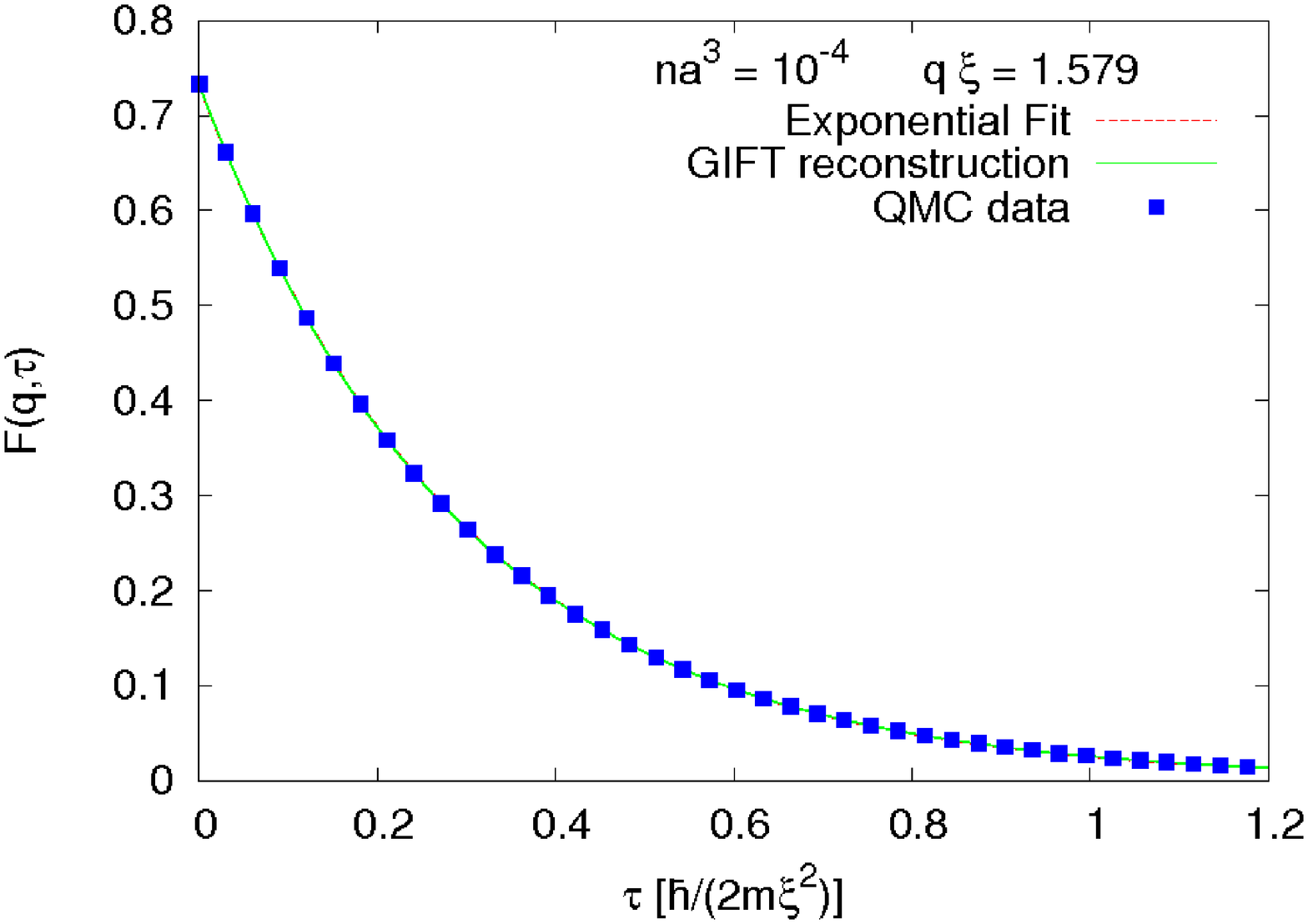}
\label{subfig3_1}
}
\subfigure[]{
\includegraphics[angle=-90,width=7.5cm]{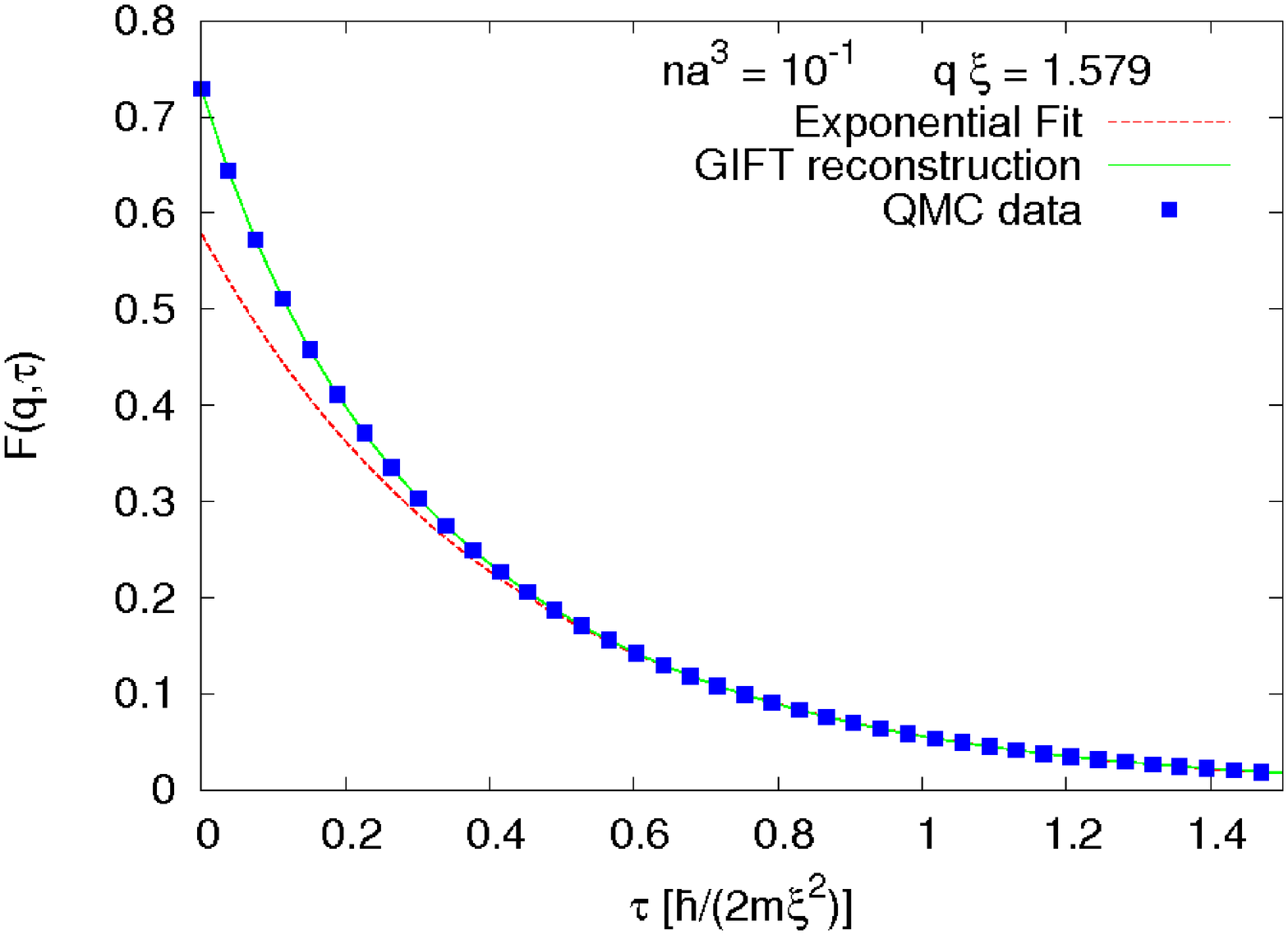}
\label{subfig3_2}
}
\caption[Optional caption for list of figures]{(color online). Scattering function $F({\bf q},\tau)$ for two different values of the interaction strength. The (green) solid line corresponds to the GIFT reconstruction of the signal, the (red) dashed line is the exponential fit to the long-time tail. In the upper panel the exponential fit coincides with and is hidden by the GIFT reconstruction.}
\label{figure3}
\end{figure}

Typical results of the imaginary-time correlation function $F({\bf q},\tau)$ obtained from QMC simulations are shown in Fig.~\ref{figure3} for two values of the gas parameter $na^3$.

\begin{figure}[t]
\includegraphics[angle=-90,width=9.0cm]{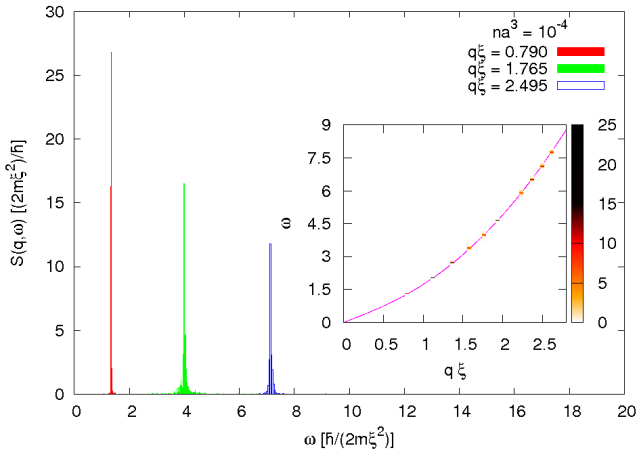}
\caption{(color online).  Dynamic structure factor at $na^3=10^{-4}$ for different values of the wavevector $q$. Inset: Color map of $S({\bf q},\omega)$ as a function of $q$. The Bogoliubov dispersion (\ref{GAS2}) is shown for comparison.}
\label{figure4}
\end{figure}

In the regime of weak correlations, one expects that mean-field theory provides an accurate description of the dynamics of the gas. Within this approach the dynamic structure factor is exhausted by a single excitation 
\begin{equation}
S({\bf q},\omega)=S({\bf q})\delta(\omega-\epsilon_{\bf q}^{\text{BOG}}/\hbar) \;,
\label{GAS1}
\end{equation}   
where 
\begin{equation}
\epsilon_{\bf q}^{\text{BOG}}=\frac{\hbar^2}{2m\xi^2}\sqrt{(q\xi)^4+2(q\xi)^2}
\label{GAS2}
\end{equation}
is the Bogoliubov spectrum written in terms of the healing length $\xi=1/\sqrt{8\pi na}$. By using Eq.~(\ref{GIFT4}), the scattering function $F({\bf q},\tau)$ is given, in this case, by the exponential function
\begin{equation}
F({\bf q},\tau)=S({\bf q}) \; e^{-\tau\epsilon_{\bf q}^{\text{BOG}}/\hbar} \;.
\label{GAS3}
\end{equation}
At $na^3=10^{-4}$, a single exponentially decaying function fits indeed very well the whole time evolution of $F({\bf q},\tau)$ for all values of ${\bf q}$ [see Fig.~\ref{subfig3_1}] and the extracted excitation energy is in excellent agreement with the Bogoliubov spectrum (\ref{GAS2}) as it is shown in Fig.~\ref{figure4}.

For larger values of the gas parameter, the long-time tail of $F({\bf q},\tau)$ can still be well fitted by an exponential function corresponding to the lowest excitation peak [see Fig.~\ref{subfig3_2}]. The short-time decay of the correlation function is instead dominated by higher energy multiphonon excitations, which cannot be described by a simple exponential law. This is clearly shown in Fig.~\ref{subfig3_2} where one has to use the function $F({\bf q},\tau)$ reconstructed from $S_{GIFT}({\bf q},\omega)$ using Eq.~(\ref{GIFT4}) to reproduce the QMC data at all times.

\begin{figure}[t]
\includegraphics[angle=-90,width=9.0cm]{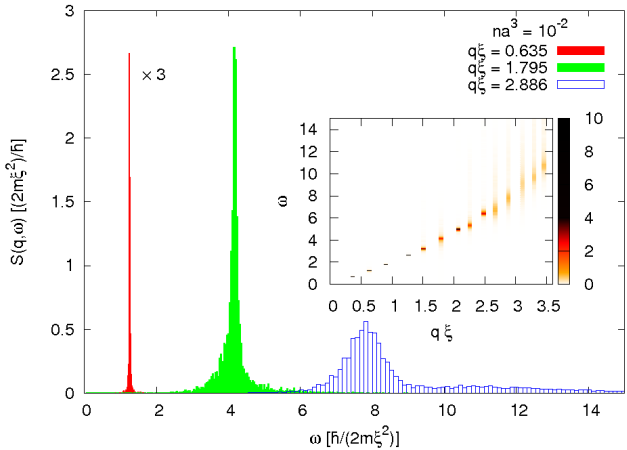}
\caption{(color online).  Dynamic structure factor at $na^3=10^{-2}$ for different values of the wavevector $q$. The signal corresponding to $q=0.635/\xi$ is rescaled by a factor 0.35. Inset: Color map of $S({\bf q},\omega)$ as a function of $q$.}
\label{figure5}
\end{figure}

\begin{figure}[t]
\includegraphics[angle=-90,width=9.0cm]{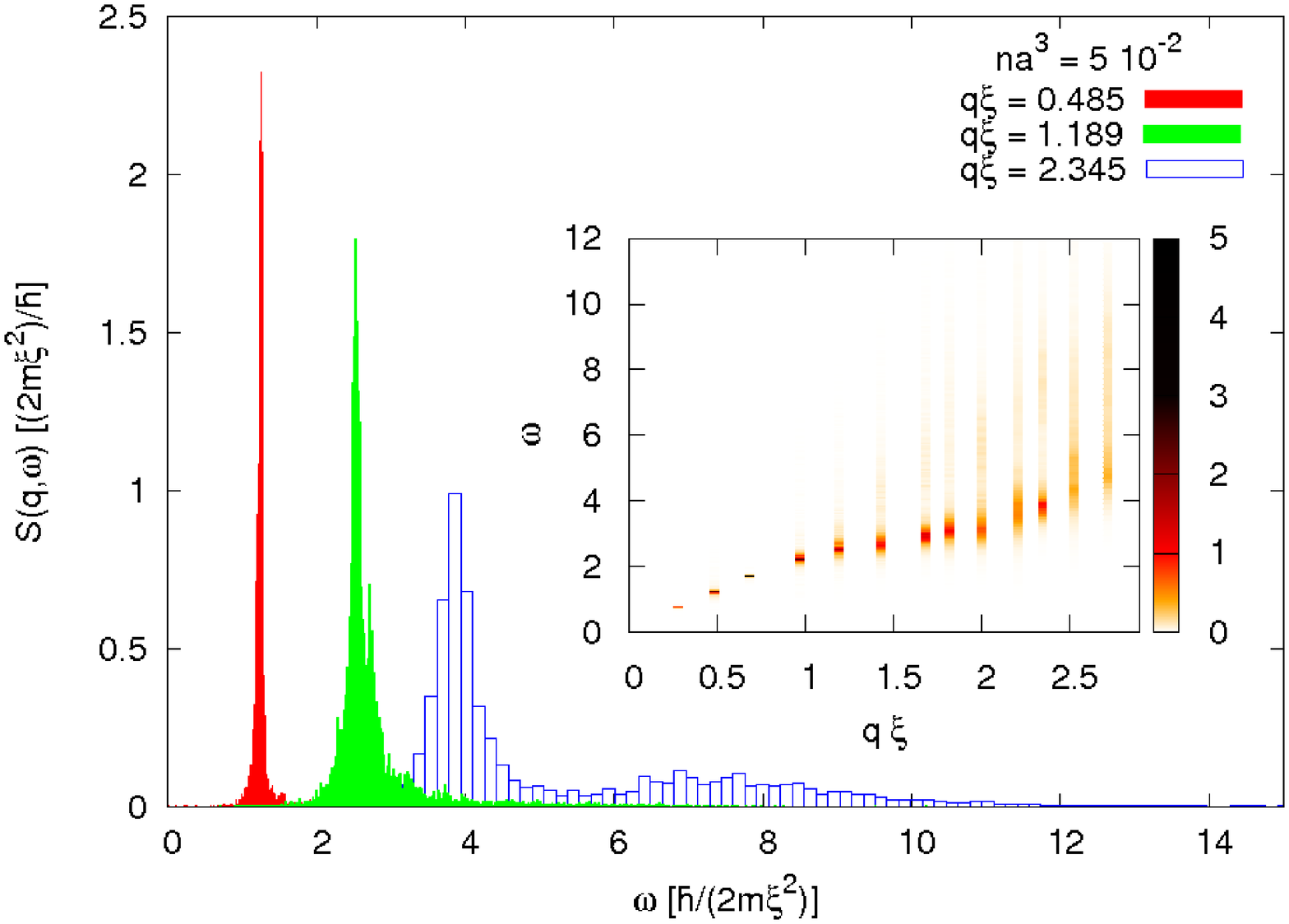}
\caption{(color online). Dynamic structure factor at $na^3=5\times10^{-2}$ for different values of the wavevector $q$. Inset: Color map of $S({\bf q},\omega)$ as a function of $q$.}
\label{figure6}
\end{figure}

\begin{figure}[t]
\includegraphics[angle=-90,width=9.0cm]{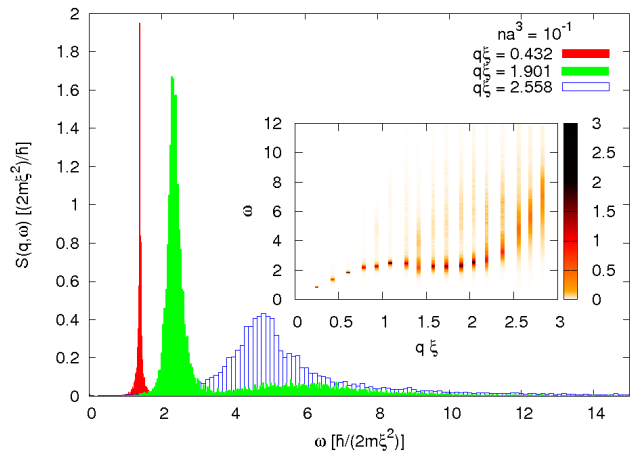}
\caption{(color online). Dynamic structure factor at $na^3=10^{-1}$ for different values of the wavevector $q$. Inset: Color map of $S({\bf q},\omega)$ as a function of $q$.}
\label{figure7}
\end{figure}

Spectra of $S({\bf q},\omega)$ for different values of $na^3$ obtained from the GIFT algorithm are shown in Figs.~\ref{figure4}-\ref{figure7}. At the smallest value of the interaction strength, $na^3=10^{-4}$ (see Fig.~\ref{figure4}), the dynamic structure factor is exhausted, for all reported values of ${\bf q}$, by a single narrow peak corresponding to the excitation of a quasiparticle with energy $\hbar\omega({\bf q})$. The dispersion of the peak with the wavevector ${\bf q}$ follows closely the Bogoliubov spectrum (\ref{GAS2}) (see inset). A small damping of the quasiparticles, Beliaev damping, was predicted in Ref.~\onlinecite{Beliaev58}, resulting in a broadening of the excitation peak. We attribute the small width of the peaks in Fig.~\ref{figure4} to the limited ability of the GIFT method to determine their exact position in the inversion procedure and not to the physical processes involved with Beliaev damping.

\begin{figure}[t]
\includegraphics[height=8.0cm,angle=-90]{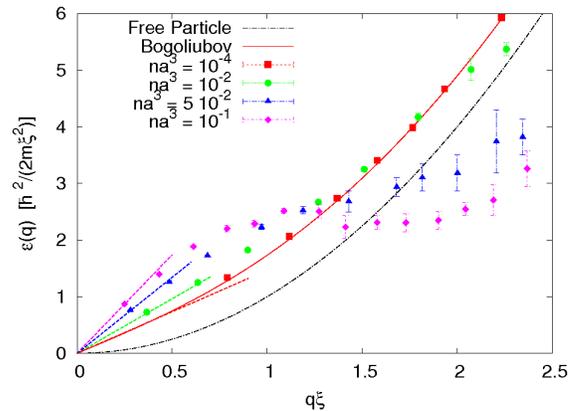}
\caption{(color online). Dispersion of the central position of the highest peak in $S({\bf q},\omega)$ for different values of the gas parameter. 
The error--bars represent the 1/2--height widths of the peaks. The solid line is the Bogoliubov prediction and the dot--dashed line corresponds to the free-particle dispersion $(q\xi)^2$. The dashed lines show the phonon dispersion $c\hbar q$, where the speed of sound $c$ is calculated from the equation of state.}
\label{figure8}
\end{figure}

The results at $na^3=10^{-2}$ are shown in Fig.~\ref{figure5}. While at the smallest wavevector $S({\bf q},\omega)$ is still given by a narrow peak centered at the phonon energy $\hbar\omega({\bf q})=c\hbar q$ where $c$ is the speed of sound (see Fig.~\ref{figure8}), at larger values of ${\bf q}$ some broadening of the peak becomes clearly visible as well as some high-frequency tail of the spectral function. These features are more evident at $na^3=5\times10^{-2}$ (see Fig.~\ref{figure6}) where, in addition, a secondary broad multiphonon peak starts to appear at large wavevectors.  We notice, from the insets of Figs.~\ref{figure5}-\ref{figure6}, that the curvature of the spectrum at small wavevectors changes from positive to negative in this interval of values of the gas parameter. A positive curvature, similar to the Bogoliubov spectrum (\ref{GAS2}), implies that long-wavelength phonons can decay into pairs of phonons by means of Beliaev processes. These decaying mechanisms are instead forbidden if the curvature of the spectrum is negative. We also notice that, at the value $na^3=5\times10^{-2}$ of the gas parameter, the dispersion of the central position of the highest peak exhibits a shoulder in the region $1.5\lesssim q\xi\lesssim2$. This shoulder develops into a visible minimum at the largest interaction strength $na^3=10^{-1}$ (see Fig.~\ref{figure7}). In the region of the minimum, $q\xi=1.901$, one can clearly distinguish one relatively narrow peak at small frequency, corresponding to the excitation of a single quasiparticle, from a broad multiphonon peak at higher frequency.

\begin{figure}[t]
\includegraphics[width=8.5cm,angle=0]{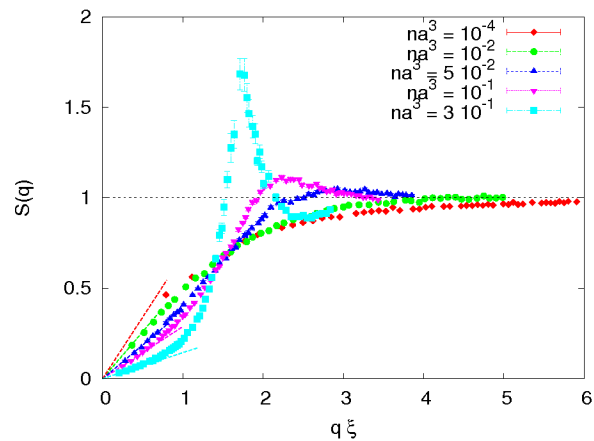}
\caption{(color online). Static structure factor $S({\bf q})$ for various densities in the gas phase. The dashed lines at small momenta correspond to the linear slope $S({\bf q})=q/(2mc)$, where $c$ is the speed of sound calculated from the equation of state.}
\label{figure9}
\end{figure}

The appearance of a minimum in the energy spectrum is connected with that of a maximum in the static structure factor $S({\bf q})$. Results of $S({\bf q})$ for the values of the gas parameter considered in this work are reported in Fig.~\ref{figure9}. At densities larger than $na^3\gtrsim0.1$, a peak starts to develop for wavevectors on the order of $2\pi n^{1/3}$, signaling the appearance of local shell structures typical of a dense gas. A similar feature in $S({\bf q})$ is exhibited by superfluid $^4$He, where the well-known roton minimum in the spectrum of excitations is occurring. Nevertheless, the description provided by the Feynman relation $\hbar\omega({\bf q}) = \hbar^2q^2/[2mS({\bf q})]$, which is easily derived from the assumption that $S({\bf q},\omega)$ is exhausted by a single quasiparticle peak, is correct only qualitatively. Our results show that multiphonon excitations become relevant at densities $na^3\gtrsim 10^{-2}$. As a consequence, the Feynman relation can provide reliable quantitative results only in the limit of weakly interacting systems.

The dispersion of the central position of the highest peak in $S({\bf q},\omega)$ is reported in Fig.~\ref{figure8} as a function of the wavevector ${\bf q}$ for different values of the gas parameter. As we already pointed out, at $na^3=10^{-4}$ the dispersion curve is in good agreement with the prediction of Bogoliubov theory [Eq.~(\ref{GAS2})] holding for dilute systems. For larger values of $na^3$ deviations start to appear both in the phonon region ($q\xi<1$), where the excitation energy is higher than $\epsilon_{\bf q}^{\text{BOG}}$, and in the single-particle region ($q\xi>1$), where $\hbar\omega({\bf q})<\epsilon_{\bf q}^{\text{BOG}}$.
The roton minimum becomes clearly visible for the largest value of the interaction strength. Our results show that no long--range attractive tail in the interaction potential is needed to observe a roton minimum in the density fluctuation spectrum. We also point out that, for all values of the gas parameter, the dispersion of the main peak in $S({\bf q},\omega)$ at the smallest wavevectors is in good agreement with the phonon dispersion $\hbar\omega({\bf q})=c\hbar q$, where $c$ is the speed of sound which we determine from the compressibility relation $mc^2=n\frac{d\mu}{dn}$, involving the chemical potential $\mu=\frac{dE}{dN}$. In practice, we calculate the value of $c$ as a function of the gas parameter by using the fit to the equation of state $E(n)$ of the HS gas found in Ref.~\onlinecite{Boronat00}. The good agreement found between the calculated excitation spectrum at small wavevectors and the phonon dispersion law determined from the equation of state in the thermodynamic limit, indicates that in our QMC simulations finite-size effects are kept under control.

The speed of sound $c$ also enters the linear slope $S({\bf q})\simeq\frac{q}{2mc}$, characterizing the static structure factor at small momenta and arising from phonon excitations. Fig.~\ref{figure9} shows that our QMC results for $S({\bf q})$ well reproduce this asymptotic law. Only at the lowest density, $na^3=10^{-4}$, the smallest values of ${\bf q}$ set by the size of the simulation box appear to be already slightly outside the phonon regime.

\subsection{Comparison with superfluid $^{\bf 4}$He}

Particular interest has to be devoted to the study of the HS system at the gas parameter $na^3=0.2138$. The HS gas at this density has been used as a reference system for the simulation of superfluid $^4$He at equilibrium in a previous work which considers the hard-wall potential as the leading part of the He-He interaction and the attractive tail as a weak perturbation\cite{Kalos74}. The value $na^3 = 0.2138$ is obtained from the experimental density of liquid $^4$He at saturated vapor pressure and from the s-wave scattering lenght of the repulsive part of the Lennard-Jones potential which models the He-He interaction. 

\begin{figure}[t]
\includegraphics[angle=-90,width=9.0cm]{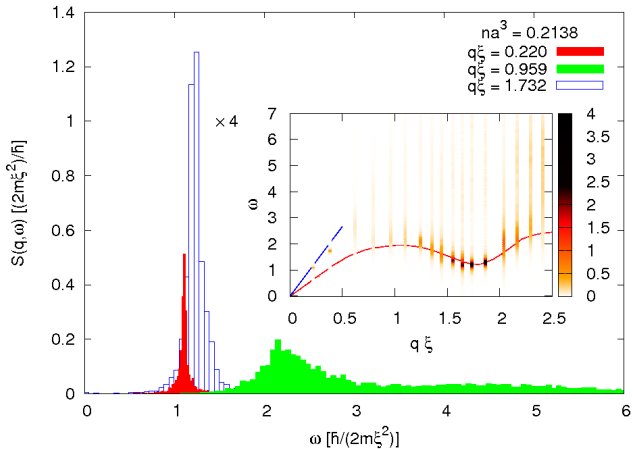}
\caption{(color online). Dynamic structure factor at $na^3= 0.2138$ for different values of the wavevector $q$. Inset: Color map of $S({\bf q},\omega)$ as a function of $q$; the red line represents the experimental dispersion of the elementary excitations in liquid $^4$He at saturated vapor pressure~\cite{Donnelly81} and the blue line represent the phonon dispersion $\omega({\bf q})=c q$, with $c$ obtained from the equation of state of a HS gas \cite{Boronat00}.}
\label{SpectralFuncHelium}
\end{figure}

Results for $S({\bf q},\omega)$ in the HS gas at $na^3=0.2138$ are showed in Fig. \ref{SpectralFuncHelium}: we see that the spectral function presents a sharp quasiparticle peak for small values of ${\bf q}$ and in the region where the excitation spectrum displays the minunum $1.5\lesssim q\xi\lesssim2$, while it is relatively broader for wavevectors $0.5\lesssim q\xi\lesssim 1.5$ between these two regimes. In the inset of Fig. \ref{SpectralFuncHelium}, we also compare the spectra obtained in our work with the dispersion curve for elementary excitations in superfluid $^4$He at low temperature, obtained from inelastic neutron scattering experiments \cite{Donnelly81}, conveniently rescaled in the units of $\xi^{-1}$ for the wavevector and of $\hbar/(2 m \xi^2)$ for the frequency.
At small ${\bf q}$, we see that the dispersion of the main peak of $S({\bf q},\omega)$ is linear, as we expect from the phonon dispersion law $\hbar\omega({\bf q})=c\hbar q$: the value for the speed of sound $c$ obtained from our result is in agreement with the one calculated from the equation of state for a HS gas \cite{Boronat00}, but it is larger than the value obtained from the experimental measurements on superfluid $^4$He, indicating, as one should expect, that the attractive part of the He-He potential plays a relevant role in determining the velocity of sound in the system. On the contrary, the dispersion of the main peak of $S({\bf q},\omega)$ is in good agreement with the experimental excitation spectrum in the roton region and for higher momenta. 

\begin{figure}[t]
\includegraphics[angle=-90,width=9.0cm]{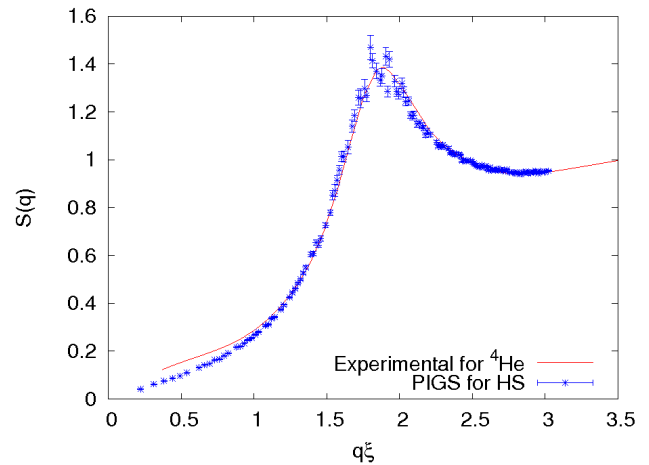}
\caption{(color online). Static structure factor $S({\bf q})$ for the HS gas at $na^3 = 0.2138$ (blue symbols). The red solid line represents the static structure factor measured in liquid $^4$He at $T = 1 \, {\rm K}$ by neutron scattering experiments \cite{Svensson80}.}
\label{StaticStructureHelium}
\end{figure}

In Fig.~\ref{StaticStructureHelium}, we show the PIGS results for the static structure factor $S({\bf q})$ of the HS gas at $na^3 = 0.2138$ and we compare these with experimental measurements in liquid $^4$He at $T = 1 \, {\rm K}$ and saturated vapor pressure\cite{Svensson80}. We can see that the agreement is excellent for $q \xi \gtrsim 1$, indicating that the HS model is able to reproduce the microscopic structure of the $^4$He system at distances comparable and smaller than the mean interparticle separation. Deviations between the PIGS results and the experimental curve are instead visible in the range $q \xi \lesssim 1$, which arise from the inability of the HS model to describe the long-range correlations among the $^4$He atoms. Thermal effects are not expected to affect the comparison since, at such low temperature, they are negligible for $q \gtrsim 0.3$~\AA$^{-1} = 0.277~\xi^{-1}$ (see Ref. ~\onlinecite{Achter69}).

In conclusion, these remarkable results indicate clearly that the structure and the density fluctuation spectrum of superfluid $^4$He, for wavevectors larger and on the order of the inverse mean interparticle distance, can be well described in terms of the hard-core repulsive potential alone.

\subsection{Solid and metastable phases}

\begin{figure}[t]
\includegraphics[angle=-90,width=9.0cm]{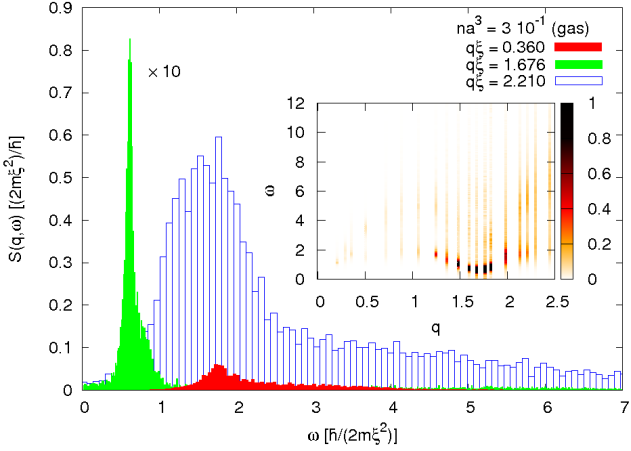}
\caption{(color online). Dynamic structure factor in the metastable gas phase at $na^3=3\times10^{-1}$ for different values of the wavevector $q$. The signal corresponding to $q=1.676/\xi$ is rescaled by a factor 0.1. Inset: Color map of $S({\bf q},\omega)$ as a function of $q$. We set to 1 the maximum value of the color scale in the contour plot, in order to show the contribuions at low spectral weight.}
\label{figure10}
\end{figure}

\begin{figure}[t]
\includegraphics[width=9.0cm]{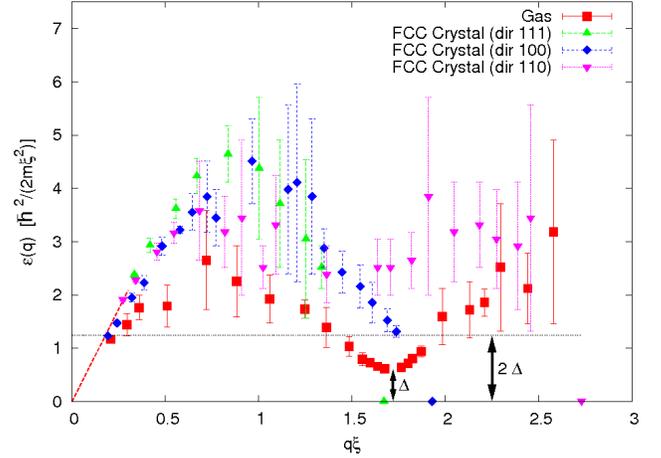}
\caption{(color online). Dispersion of the central position of the highest peak in $S({\bf q},\omega)$ at $na^3=3\times10^{-1}$ in the metastable gas phase (red squares) and in the fcc crystal phase: direction (1,1,1) green up triangles, direction (1,0,0) blue diamonds, direction (1,1,0) magenta down triangles. Similarly to Fig.~\ref{figure8}, the dashed line shows the phonon dispersion $\hbar cq$ in the gas phase. The horizontal dotted line corresponds to twice the roton gap $\Delta$.}
\label{figure11}
\end{figure}

For values of the gas parameter larger than the melting density, $n_ma^3=0.288(1)$, the thermodynamically stable phase is the crystal, while the gas state can only survive as a metastable phase (see Fig.~\ref{figure1}). In Fig.~\ref{figure10} we show the dynamic structure factor in the metastable gas phase at $na^3=3\times10^{-1}$.  We notice that the distribution of spectral weight is in general very broad and extends to relatively high frequencies. Only in the roton region a narrow quasiparticle peak is present accompanied by multiphonon excitations at higher frequencies. More details about the spectrum of excitations in the metastable gas phase can be extracted from Fig.~\ref{figure11} where we plot the dispersion of the central position of the highest peak as a function of the wavevector ${\bf q}$. The roton minimum is clearly evident at energies significantly lower than in the gas phase at $na^3=10^{-1}$ (see Fig.~\ref{figure8}). The small error bars indicate that the peak is indeed well defined in the region around the roton minimum corresponding to the energy gap $\Delta$. The width of the peak increases dramatically as soon as the excitation energy is above the threshold $2\Delta$ in agreement with the theoretical explanation in terms of two-roton quasiparticle decay processes. Similarly to Fig.~\ref{figure8}, the points corresponding to the smallest wavevectors available in our simulation box $q=\frac{2\pi}{V^{1/3}}$ agree with the phonon dispersion $\hbar\omega({\bf q})=\hbar c q$, where the speed of sound $c$ is estimated from the equation of state.

The wavevector dispersions of the highest peaks of $S({\bf q},\omega)$ in the fcc crystal phase at $na^3=3\times10^{-1}$ are shown in Fig.~\ref{figure11} along three independent spatial directions. 
The three spectra agree at small momenta, where they all converge to the energy $\hbar\omega({\bf q})=c_L\hbar q$ of longitudinal phonons propagating with the speed $c_L$, while at larger values of ${\bf q}$ they differ significantly and the spectral intensities associated to them have a vanishing energy at different points corresponding to the wavevectors of the reciprocal lattice. Here, the spectral function is exhausted by the elastic peak at $\omega=0$ and the static structure factor $S({\bf q})$ diverges with the number $N$ of particles in the system. We notice that, when the wavevector lies in the (1,1,1) direction and explores the diagonal of the elementary cell, the smallest non-zero wavevector of the reciprocal lattice is at $q\simeq1.7/\xi$.
The roton minimum in the metastable gas phase at the same value of the density of the solid is found very near this value of ${\bf q}$.
The broad signals obtained in the central region of the Brillouin zone presumably arise from the fact that the energies of the three branches are not well separated from the higher phonon bands and the GIFT method cannot resolve the different contributions to the dynamic structure factor.

\section{Conclusions}

In conclusion, we investigated the dynamic structure factor and the elementary excitations of a system of bosonic hard spheres at $T=0$ in the gas phase ranging from the dilute to the dense regime, up to the solid phase reached at very high density. The numerical method used is one of the most reliable among the ones presently available and it consists of the implementation of the GIFT algorithm on top of results of imaginary-time correlation functions obtained via QMC simulations. In this way, as the density of the gas increases, we follow the change of the dynamic structure factor from a narrow peaked feature whose dispersion closely agrees with Bogoliubov spectrum, to a superposition of a relatively narrow quasiparticle peak and a broader, high frequency  multiphonon contribution where the quasiparticle energy exhibits the typical phonon--roton dispersion similar to superfluid $^4$He. We find that the roton minimum emerges in the spectrum within the range of gas parameters $0.05<na^3<0.1$. No long--range attractive tail in the interaction potential is thus needed to observe a well defined roton in the density fluctuation spectrum.
We studied also the HS model with a gas parameter $na^3=0.2138$, corresponding to superfluid $^4$He at equilibrium density; remarkably, the energy--momentum dispersion relation in the roton region and at higher momenta turns out to be in good agreement with the measured experimental spectrum of superfluid $^4$He. This suggests that in strongly interacting systems characterized by a hard-core interatomic potential at short distances, the HS model is able to capture, at least semi--quantitatively, the density fluctuation spectrum for wavevectors reciprocal to interparticle and lower distances.
Above the freezing density in the metastable gas phase the roton minimum deepens and it occurs near a value of ${\bf q}$ corresponding to the smallest wavevector of the reciprocal lattice where, in the solid phase, the strength of the dynamic structure factor vanishes.

\section*{Acknowledgements}
We would like to acknowledge L.P. Pitaevskii and L. Reatto for useful discussions. This work has been supported by Regione Lombardia and CINECA Consortium through a LISA Initiative (Laboratory for Interdisciplinary Advanced Simulation) 2012 grant [http://www.hpc.cineca.it/services/lisa], by ERC through the QGBE grant and by Provincia Autonoma di Trento. Partial support by the Italian MIUR under Contract Cofin-2009 ``Quantum gases beyond equilibrium'' is also acknowledged. We thank the Aurora-Science project (funded by PAT and INFN) for allocating part of the computing resources for this work and for technical support.

\end{document}